\documentclass[aip,11pt,a4paper]{revtex4}
\usepackage{graphicx}
\usepackage{hhline}
\usepackage{float}
\usepackage{bm}
\usepackage{color}
\usepackage[normalem]{ulem}

\usepackage{latexsym,amsmath, amscd, amssymb} 
\usepackage[caption=false]{subfig}
\usepackage{graphicx}
\usepackage{epstopdf}
\usepackage{multirow}
\usepackage{url}
\usepackage{rotating}
\usepackage{placeins}
\usepackage{hyperref}
\hypersetup{colorlinks, citecolor=red, filecolor=yellow,linkcolor=blue,urlcolor=cyan,linktocpage}

\draft 

\begin{document}


\title{Robust and efficient configurational molecular sampling via Langevin Dynamics} 



\author{Benedict Leimkuhler}
\author{Charles Matthews}
\email[Send correspondance to:]{charles.matthews@ed.ac.uk}
\thanks{School of Mathematics and Maxwell Institute of Mathematical Sciences, James Clerk Maxwell Building, Kings Buildings, University of Edinburgh, Edinburgh, EH9 3JZ, UK}
\affiliation{School of Mathematics and Maxwell Institute of Mathematical Sciences, James Clerk Maxwell Building, Kings Buildings, University of Edinburgh, Edinburgh, EH9 3JZ, UK}


\date{\today}

\begin{abstract}
A wide variety of numerical methods are evaluated and compared for solving the stochastic differential equations encountered in molecular dynamics.  The methods are based on the application of  deterministic impulses, drifts, and Brownian motions in some combination.  The Baker-Campbell-Hausdorff expansion is used  to study sampling accuracy following recent work by the authors, which allows determination of the stepsize-dependent bias in configurational averaging.  For harmonic oscillators, configurational averaging is exact for certain schemes, which may result in improved performance in the modelling of biomolecules where bond stretches play a prominent role.  For general systems, an optimal method can be identified that has very low bias compared to alternatives.   In simulations of the alanine dipeptide reported here (both solvated and unsolvated), higher accuracy is obtained without  loss of computational efficiency, while allowing large timestep, and with no impairment of the conformational exploration rate (the effective diffusion rate observed in simulation).  The optimal scheme is a uniformly better performing algorithm for molecular sampling, with overall efficiency improvements of 25\% or more in practical timestep size achievable in vacuum, and with reductions in the error of configurational averages of a factor of ten or more attainable in solvated simulations at large timestep.

\end{abstract}


\maketitle 

\section{Introduction }
%
%
%

\begin{figure}
\includegraphics[trim = 0mm 0mm 0mm 180mm, clip=true,width=4in]{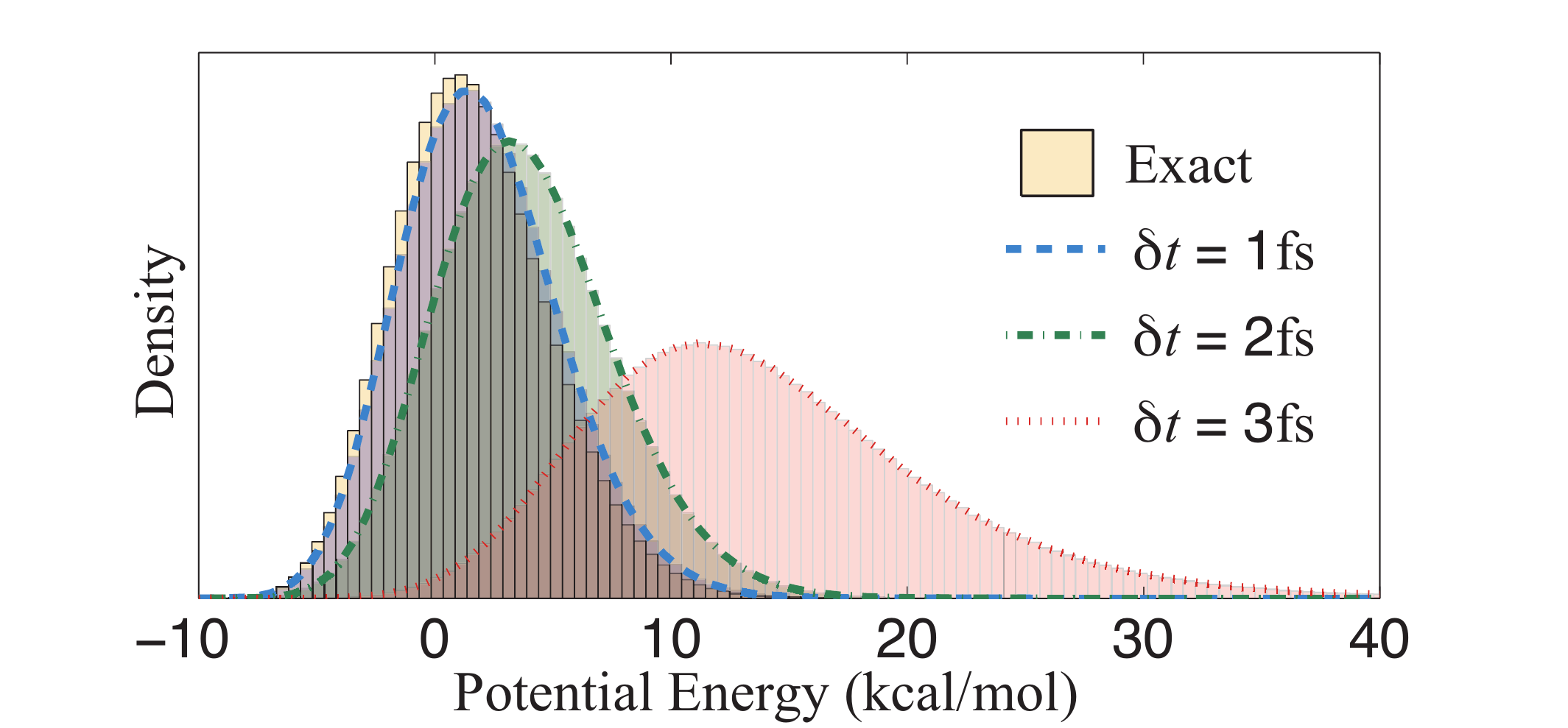}
\caption{The computed potential energy distribution is shown for the method of Br\"{u}nger, Brooks and Karplus applied to a single alanine dipeptide protein at 300K in a vacuum using the CHARMM22 forcefield; the energy distribution becomes distorted as the step size increases.
} \label{stepsizeeffect}
\end{figure}

One of the major challenges in understanding matter at the molecular scale is the problem of thermodynamic sampling: the calculation of averages with respect to the canonical (Gibbs-Boltzmann) distribution.   In many cases the aim is to sample configurational quantities only, and this is the focus of this article.  Given the classical molecular potential energy function $U:\mathbb{R}^{3N} \rightarrow \mathbb{R}$, the configurational canonical density  is
\[
\bar{\rho}_\beta(q) = Z^{-1} e^{-\beta U(q)},
\]
where $\beta^{-1} = k_BT$ where $k_B$ is Boltzmann's constant, $T$ is temperature, and $Z$ is  a normalization constant so that $\bar{\rho}_{\beta}$ has unit integral over the entire configuration space.      In using molecular dynamics to sample the phase space according to the canonical distribution, the formulation employed may not be ergodic (meaning that it may not sample the entire phase space) and, moreover,  the design of the time-discretization methods typically distorts the equilibrium distribution.  Using a stochastic differential equation model such as Langevin dynamics, which introduces random perturbations into each force component, we can overcome the first of these problems, as the formulation is well known to be ergodic.   As an illustration of the effect of step size error, see Figure \ref{stepsizeeffect}, where the potential energy in simulations of alanine dipeptide is shown to be corrupted by a popular Langevin discretization.

Given that the majority of computational work in any MD algorithm lies in the force calculation, most of the existing methods in common use have been designed to require only one force evaluation per timestep.  For timestepping methods that accurately sample the canonical distribution, the available timescales for simulation are restricted by the problem itself (e.g. the heights of barriers, or entropic properties of the landscape).   In designing a new molecular dynamics algorithm the goal is to enlarge the usable timestep in order to allow finite trajectories to access a larger portion of the phase space.
The drawback of working in the high-timestep regime is that for long-time simulations the computed probability distribution is a perturbation of $\bar{\rho}_\beta$, dependent on the step size, leading to a distortion in calculated averages. The simulator must choose the step size sufficiently small enough to avoid corruption in averages, but still large enough to ensure a thorough exploration of configuration space. 

The potential energy function in molecular dynamics determines the maximum allowable step size.  For a harmonic oscillator with frequency $\omega$ the Verlet method has a stability restriction of $\delta t\leq 2/\omega$ \cite{AllenTildesley}.  Most numerical methods (including ones constructed for Langevin dynamics) suffer from a similar limitation in the maximum timestep size driven by the presence of stiff oscillatory solution components.  However, well before reaching the stability threshold, averages may be severely corrupted, introducing artificial--and, often, severe--step size restriction.     By removing or reducing this bias, it becomes possible to substantially increase the timestep, with a direct impact on the efficiency of simulation.   Given the explosion in the use of molecular dynamics in chemistry, physics, engineering and biology, it is worth noting that where molecular dynamics is used for round-the-clock configurational sampling calculations, a quantifiable improvement in method efficiency (or timestep size) directly translates to a reduction in machine costs, a reduction in energy costs, and, often, a reduction in delays to publication.

\begin{figure}[hbt]
\begin{center}
\includegraphics[width=5in]{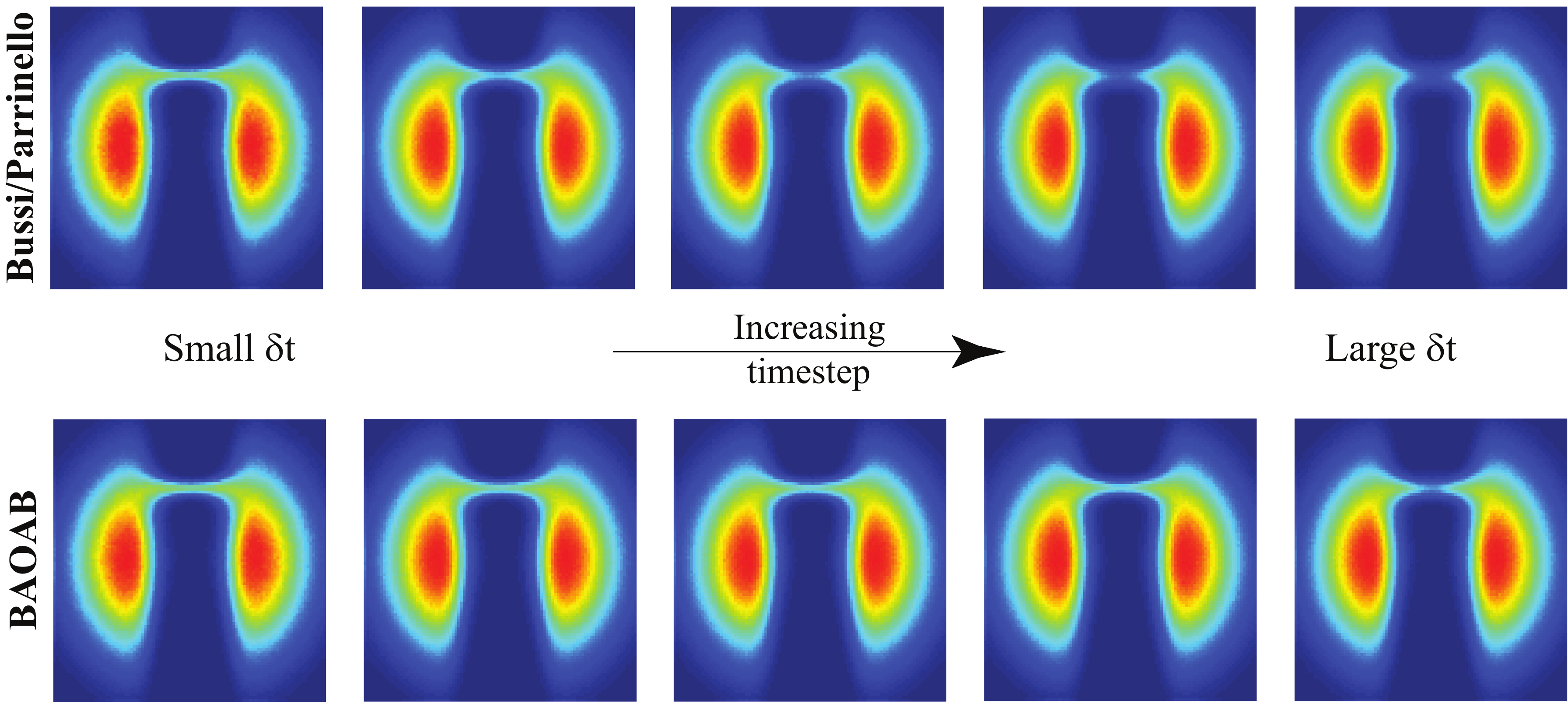}
\caption{The configurational density function $\bar{\rho}_\beta(q)$ is shown for a planar two-basin potential, computed using  two different Langevin dynamics methods at various (stable) timesteps increasing from left to right. The color indicates the value of the computed probability density: from high (red) to low (blue) over the unit square. The methods have the same computational cost (in terms of force evaluations), but give  different results at large timesteps. Details on this computation are given in Appendix \ref{app_fig2}.
\label{lungs}}
\end{center}
\end{figure}

With regard to the error in averages, it is usually assumed that the error due to having insufficient samples dominates the timestep-dependent discretization error, but this is not typically the case at large step size, as we demonstrate in numerical experiments (see Section V).
To dramatize the role of discretization error, we show in Figure \ref{lungs} the example of the configurational distribution for a simple two-basin model solved using two different numerical methods.  The figure illustrates that where step size error is substantial, crucial features of the landscape such as the heights of free energy barriers,  may be completely altered.  Moreover, it is entirely possible for the relative heights of different barriers to be altered in such a way that one transition becomes more prevalent than another.

Theoretical analysis of the error in the invariant distribution can be performed for harmonic oscillators without great difficulty (see Section III), but it can also be carried out for general nonlinear problems.  This is most straightforward in the case of splitting methods. In this article we draw on principles of geometric integration, building on our understanding of splitting methods from the deterministic setting \cite{LeRe2005}.
Splitting methods for Langevin dynamics have been considered in the past \cite{Shardlow,Athenes2004,french,BouOwh,SPV,LI,LeRoSt2012,SiChCr} but a wide variety of schemes can be constructed by splitting and until now the rational basis for selecting one scheme over another has been absent.    Drawing on the work of Talay and Tubaro \cite{TalayTubaro}, we have studied the generator of the numerical method directly by examining expansions for the invariant measure of the Langevin dynamics scheme \cite{LeMa2012}; this investigation lead to the concept of the
associated density $\hat{\rho}$ of the numerical method:
\begin{equation}
\hat{\rho}(q,p) \propto \exp \left(-\beta \left[H(q,p) + \delta t^2 f_2(q,p) + \delta t^4 f_4(q,p) + \ldots \right] \right).
\label{rhohat}
\end{equation}
This expansion allows different methods to be compared on a rational basis (as concerns the effect of discretization error).   In the past, for deterministic methods, this type of analysis has also been used for the correction of averages \cite{BoLe2007,Davidchack}.  In typical cases which would be relevant for molecular simulation, the error introduced in averages using such methods would be second order in the timestep (i.e. would go to zero quadratically as the step size is reduced).

For  a particular ordering of the building blocks of the numerical method, a ``superconvergence'' (cancellation) property can be obtained in the high friction limit, meaning that in fully resolved molecular dynamics simulations and after integrating out with respect to momenta,  the leading term in the expansion vanishes \cite{LeMa2012}.  This theoretical convergence result was until now only studied for  relatively simple model problems.  Moreover the crucial question of the step size (stability) threshold of the different schemes (as well as the overall efficiency of the various methods)   cannot be addressed using the asymptotic technique since it provides information only about the small step size limit ($\delta t\rightarrow 0$).

Another important issue raised by practitioners concerns the fact that such a superconvergent method, relying on large friction, might not be useful in realistic settings since it is known that large friction can reduce the diffusion rate.  The problem is complicated by a number of issues: both friction coefficient and step size affect the long-term averaging error differently for different methods, and the friction coefficient (and, in principle, the step size) may affect the diffusion rates differently for different methods.  Performance is further dependent on the type of problem under study.   Thus there is a need for careful study of the methods in the context of systems relevant for molecular dynamics (for example, containing both steep potentials such as Lennard-Jones and stiff bonds).  In this article we address both issues: we consider large step size and modest values of the friction coefficient, using numerical experiments to carefully examine the relative performance of a large number of different methods.

In recent years, there has been widespread interest in multiscale methods for enhanced sampling \cite{milestoning-Elber,IzSwPa2010,BoChDeGe2002,MeDiHoSc2006} and such methods likely offer the best approach to bridging the timescale gap.  We observe that work on enhanced numerical schemes for molecular dynamics remains essential as it plays a crucial underpinning
role in all the enhanced sampling approaches.   Improved trajectory generation efficiency (e.g. allowing the use of a larger basic timestep in simulation) thus has a knock-on effect on the efficiency of all the methods that rely on such trajectories.  While relative improvements of a few percentages in efficiency can already warrant a minor change in software implementation, our analysis points to a more dramatic (even {\em qualitative}) difference among various methods leading to prospects for much greater efficiencies by selecting a suitable method.  These observations are verified in model biomolecular simulations.

Hybrid Monte-Carlo \cite{HMC,FreeEnergy}, and other schemes based on Metropolis correction \cite{BouRabeeVandenEijnden}, are not discussed here, although these could be used in conjunction with several of the methods implemented.  The improvement in thermodynamic sampling obtained through the use of more accurate Langevin integrators may, in some cases, provide an alternative to Metropolis-based correction in the practical setting. All methods under discussion require one force evaluation per iteration, and hence have practically of the same computational cost.

This article proceeds as follows. In Section II we introduce Langevin dynamics in the context of configurational sampling and describe our method for examining the  long-time behavior of averages under discretization of the stochastic differential equations (SDEs).   Section III discusses the harmonic model problem, showing that for some particular schemes, the configurational sampling can be exact; this has implications for molecular simulations involving stiff harmonic bonds.   Section IV addresses the errors obtained from computed averages in more general systems. 
Section V contains numerical experiments comparing various methods, both for one degree of freedom systems and for solvated and unsolvated alanine dipeptide, through implementation of the schemes in a version of NAMD \cite{namdlite2007}.   It is our contention that the numerical experiments of Section V
provide strong evidence for rejecting many of the schemes in common use for stochastic molecular dynamics and  favor the optimal BAOAB scheme of  \cite{LeMa2012}.

\section{Background }
In this article we focus on Langevin dynamics,
\begin{eqnarray}
{\rm d} q& = & M^{-1} p \, {\rm d}t, \label{lang-1}\\
{\rm d} p& = & -\nabla U(q) \, {\rm d}t -\gamma p \, {\rm d}t + \sigma M^{1/2} \, {\rm d}W, \label{lang-2}
\end{eqnarray}
where $q,p \in \mathbb{R}^{3N}$ are vectors of instantaneous position and momenta respectively, $W=W(t)$ is a vector of $3N$ independent Wiener processes, $\gamma>0$ is a free (scalar) parameter and $M$ is a constant diagonal mass matrix. By choosing $\sigma = \sqrt{2\gamma \beta^{-1}}$ it is possible to show that the unique probability distribution sampled by the dynamics is the canonical (Gibbs-Boltzmann) density, defined as
\begin{equation}
\rho_{\beta}(q,p) = \Omega^{-1} e^{-\beta H(q,p)}, \label{BGmeas}
\end{equation}
for total system energy (Hamiltonian) $H(q,p)=p^TM^{-1}p/2 + U(q)$, and normalization constant $\Omega^{-1}$ ensuring the integral is unity. We consider numerical methods designed to integrate (\ref{lang-1}--\ref{lang-2}), primarily for the purpose of generating trajectories that sample $\rho_\beta.$ Such trajectories are often used as a means for calculating expectations of functions purely of the position $q$, and as such the dynamical fidelity of computed trajectories is of minor importance compared to the behavior of averages in the large-time limit. For such an observable $\phi$, we write the expectation as
\[
 \mathbb{E} \left[ \phi(q)  \right] = \Omega^{-1} \int \int \phi(q) \rho_\beta(q,p) \, {\rm d} q \,  {\rm d}p = Z^{-1} \int \phi(q) \bar{\rho}_\beta(q) \, {\rm d} q = \lim_{T \rightarrow \infty} T^{-1} \int_0^T \phi(q(t)) {\rm d} t,
\]
where the ergodicity of Langevin dynamics ensures a sampling of the desired probability distribution, and hence the ability to equate the long-time average along a trajectory with the corresponding spatial average. The challenge comes in integrating (\ref{lang-1}--\ref{lang-2}) effectively, and with minimal computational cost.

Given a general potential energy function $U(q)$, we cannot integrate exactly and must evolve the dynamics by discretizing in time. Advancing the state requires the use of a numerical method which aims to approximate the exact evolution.  A distribution of initial conditions $\rho$ propagated using a second-order numerical method will evolve according to the equation
\[
\frac{\partial \rho}{\partial t}= \hat{\cal L}^* \rho,
\]
where $\hat{\cal L}^*$ may be expressed in the series expansion
\begin{equation}
\hat{{\cal L}}^* = {\cal L}^*_\text{LD} + \delta t^2 {\cal L}^*_2 + {\cal O}(\delta t^{3}), \label{lhat}
\end{equation}
where ${\cal L}^*_\text{LD}$ is the operator associated to the exact propagation under Langevin dynamics and $\delta t$ is the step size. The invariant (long-time) distribution sampled by the scheme can in principle be obtained by solving the partial differential equation (PDE) $\hat{{\cal L}}^* \hat{\rho} = 0$, assuming that the perturbed operator $\hat{{\cal L}}^*$ is known.

Splitting methods\cite{Shardlow,french,SiChCr} offer a simple way of integrating the Langevin dynamics equations; the right hand side of (\ref{lang-1}--\ref{lang-2}) is divided into pieces, eg. $\dot{z} = f = f_1 + f_2$, with each piece is solved exactly in sequence. 
Recent work \cite{LeMa2012} has shown that numerical methods derived from additive splittings of the vector field enable relatively simple computation of a method's characteristic operator. The order of integration, and the choice of the splitting will define the method. 

For example, one may break Langevin dynamics into three pieces:
\begin{equation}
\left [ \!\!\begin{array}{c} {\rm d} q\\ {\rm d} p\end{array}\!\! \right ]=
\underbrace{\left [ \!\!\begin{array}{c} M^{-1} p \\0\end{array}\!\! \right ]{\rm d}t}_{\rm A} +
\underbrace{\left [\!\! \begin{array}{c} 0 \\-\nabla U(q)\end{array}\!\! \right ]{\rm d}t}_{\rm B} +
\underbrace{\left [\!\! \begin{array}{c} 0\\ -\gamma p{\rm d}t +\sigma M^{1/2}{\rm d} W \end{array} \!\!\right ],}_{\rm O}
\label{split1}
\end{equation}
which are labelled $A$, $B$ and $O$. Each of the three pieces may be solved ``exactly'': $A$ and $B$ correspond to a linear drift and kick when taken individually, while the $O$ piece is associated to  an Ornstein-Uhlenbeck equation with ``exact'' solution
\begin{align*}
q(t) &= q(0), \\
p(t) &= e^{-\gamma t} p(0) + \frac{\sigma}{\sqrt{2 \gamma}} \sqrt{1 - e^{-2\gamma t} } M^{1/2} R_t,
\end{align*}
where $R_t \sim {\cal N}(0,1)$ is a vector of uncorrelated noise processes.  (By ``exact'' we mean that this random map generates the probability distribution $\rho(q,p,t)$ defined by the solutions of the Ornstein-Uhlenbeck equation.)

Given the pieces of the splitting we code a method by giving the sequence of integration, from left to right.   The string ``ABO'' represents the method obtained by solving first the ``A'' part for a timestep $\delta t$, then the ``B'' part, and finally the ``O'' part of the system.   Where a symbol is repeated, as in ``BAOAB,'' there could be ambiguity in this representation but we will assume that the method is symmetric so that all ``A'' and ``B'' parts  in BAOAB are integrated for a half timestep.    Additionally, the methods of Bussi and Parrinello \cite{bussiparrinello} (OBABO) as well as gla-1 (BAO) and gla-2 (BABO) of Bou-Rabee and Owhadi \cite{BouOwh}, are equivalent to splitting methods using these pieces.

An alternate splitting formulation\cite{SPV}
\begin{equation}
\left [ \!\!\begin{array}{c} {\rm d} q\\ {\rm d} p\end{array}\!\! \right ]=
\underbrace{\left [ \!\!\begin{array}{c} M^{-1} p \\0\end{array}\!\! \right ]{\rm d}t }_{\rm A} +
\underbrace{\left [\!\! \begin{array}{c} 0 \\-\nabla U(q){\rm d}t -\gamma p{\rm d}t +\sigma M^{1/2}{\rm d} W \end{array} \!\!\right ],}_{\rm S}
\label{split2}
\end{equation}
has been used to define two methods: stochastic position verlet (ASA) and stochastic velocity verlet (SAS).

A technique is outlined in Section IV to calculate the operators $\hat{{\cal L}}^*$ of such splitting methods. For methods not derived from splitting the vector field, it can be more difficult to obtain their operators and examine their behavior. We compare the configurational sampling for a number of popular schemes in this article, not limiting the scope only to splitting methods.

We will consider both ABOBA and BAOAB methods\cite{LeMa2012}, the Bussi/Parrinello method\cite{bussiparrinello}, as well as the Stochastic Position Verlet method\cite{SPV}.
The Langevin Impulse (LI) method\cite{LI}, the BBK method \cite{BBK} and the method of van Gunsteren and Berendsen (VGB)\cite{vgb88} will also be compared, as these are frequently found in commercial software packages (for example, NAMD and GROMACS).
Two first-order methods, Ermak-McCammon (EM)\cite{ErMc} and Ermak-Buckholtz (EB)\cite{ErBu} will also be considered in numerical experiments, for completeness, with each method described in Appendix \ref{app_methods}.

\section{Performance of Langevin algorithms applied to the harmonic oscillator.}
We begin by considering the harmonic model problem.   Harmonic oscillators are useful in the molecular simulation setting not only because they allow analytical determination of effective distributions, but also because they can be seen to be relevant to understanding the timestep limiting features of models for crystalline solids and biomolecules.

The one-dimensional harmonic oscillator $U(q) = K q^2/2$, $q\in\mathbb{R}$ and $K>0$, is a standard test case for Langevin dynamics numerical methods, as many issues of stability and timestep in molecular dynamics simulations arise due to harmonic potentials used to model covalent bonds. For such a simple system we may explicitly write one iteration of a general numerical method evolving the dynamics as \cite{BuLy2009}
\[
\left [ \begin{array}{c} q_{n+1}\\p_{n+1}\end{array} \right ] \leftarrow \Psi \left [ \begin{array}{c} q_{n}\\p_{n}\end{array}  \right ] + \mu_n,
 \]
where $\Psi$ is a matrix depending only on the step size $\delta t$, the friction coefficient $\gamma$, the particle mass $M$ and the spring constant $K$, while $\mu_n$ is a vector of stochastic processes.

Let $\Psi=(\psi_{ij})$  and denote the components of $\mu_n$  by $\mu_{n,j}$ where $i,j \in \{1,2\}$. Taking products of the update equations, we obtain
\begin{align}
q_{n+1}^2 &= \psi_{11}^2 q_n^2 + \psi_{12}^2 p_n^2 + \mu_{n,1}^2 + 2 \psi_{11}\psi_{12}q_np_n + 2\psi_{11}\mu_{n,1} q_n + 2\psi_{12}\mu_{n,1} p_n, \label{ho_qq} \\
p_{n+1}^2 &= \psi_{21}^2 q_n^2 + \psi_{22}^2 p_n^2 + \mu_{n,2}^2 + 2\psi_{21}\psi_{22} q_n p_n + 2\psi_{21}\mu_{n,2} q_n + 2\psi_{22}\mu_{n,2} p_n ,\label{ho_pp} \\
q_{n+1}p_{n+1} &= \psi_{11}\psi_{21} q_n^2 + \psi_{12}\psi_{22} p_n^2 + \mu_{n,1}\mu_{n,2} \nonumber\\
& \quad + (\psi_{11}\psi_{22} + \psi_{12}\psi_{21}) q_n p_n + (\psi_{11} \mu_{n,2} + \psi_{21} \mu_{n,1}) q_n + (\psi_{22} \mu_{n,1} + \psi_{12} \mu_{n,2}) p_n. \label{ho_qp}
\end{align}
We then take expectations of (\ref{ho_qq}-\ref{ho_qp}) in the limit $n\rightarrow \infty$, giving simultaneous equations
\begin{align}
\langle  q^2 \rangle  &= \psi_{11}^2 \langle q^2\rangle  + \psi_{12}^2 \langle  p^2 \rangle  + \langle \hat{\mu}_{1}^2\rangle  + 2\psi_{11}\psi_{12} \langle q p \rangle  + 2\psi_{11} \langle  \hat{\mu}_{1} q \rangle  + 2\psi_{12}\langle \hat{\mu}_{1} p \rangle  , \label{ho_exp1} \\
\langle p^2 \rangle  &= \psi_{21}^2 \langle  q^2 \rangle  + \psi_{22}^2 \langle  p^2 \rangle  + \langle \hat{\mu}_{2}^2 \rangle  + 2\psi_{21}\psi_{22} \langle  qp \rangle   + 2\psi_{21} \langle \hat{\mu}_{2} q \rangle  + 2\psi_{22} \langle \hat{\mu}_{2} p \rangle , \\
\langle qp \rangle  &= \psi_{11}\psi_{21} \langle  q^2 \rangle  +\psi_{12}\psi_{22} \langle  p^2 \rangle  + \langle \hat{\mu}_{1}\hat{\mu}_{2}\rangle  + (\psi_{11}\psi_{22} + \psi_{12}\psi_{21}) \langle  q p \rangle \nonumber \\
& \quad +\psi_{11} \langle \hat{\mu}_{2}q\rangle  + \psi_{21} \langle \hat{\mu}_{1}q\rangle  + \psi_{22}\langle  \hat{\mu}_{1}p\rangle  + \psi_{12} \langle \hat{\mu}_{2}p\rangle , \label{ho_exp2}
\end{align}
where we use notation $\langle  x \rangle  = \mathbb{E}[x_n]$ for $x \in \{q,p\}$, and  $\langle  \hat{\mu}_i \rangle  = \mathbb{E}[\mu_{n,i}]$. The value of $\langle \hat{\mu}_i x\rangle = \mathbb{E}[\mu_{n,i} \, x_n]$ will ultimately depend on the ``memory'' of a scheme's stochastic process ${\mu}_n$, and can be found by computing $x_{n+1}\, \mu_{n+1,i}$ and taking expectations, yielding an expression involving $\mathbb{E}[ \mu_{n,i} \mu_{n-1,i} ]$ .

We can hence solve the linear system (\ref{ho_exp1}-\ref{ho_exp2}) to find the error in long-time averages for a method, and its behavior under changes in $\delta t$ and $\gamma$, relative to the spring constant $K$, by comparing the numerical and  analytic averages, the latter given as
\[
\left [ \begin{array}{c} \langle  q^2 \rangle ^*\\ \langle  p^2 \rangle ^*\\\langle  qp \rangle ^* \\ \end{array}  \right ] = \left [ \begin{array}{c} K^{-1} \beta^{-1} \\ M \beta^{-1} \\ 0 \\ \end{array}  \right ].
\]
For the BAOAB and ABOBA methods, the numerical values (in the long-time limit) are
\[
\left [ \begin{array}{c} \langle  q^2 \rangle ^\text{(BAOAB)} \\ \langle  p^2 \rangle ^\text{(BAOAB)} \\  \langle  qp \rangle ^\text{(BAOAB)}  \\ \end{array}  \right ] = \left [ \begin{array}{c} K^{-1} \beta^{-1} \\ M \beta^{-1} \left(1-\frac{\delta t^2 K}{4M}\right) \\ 0 \\ \end{array}  \right ], \qquad \left [ \begin{array}{c} \langle  q^2 \rangle ^\text{(ABOBA)} \\ \langle  p^2 \rangle ^\text{(ABOBA)}  \\  \langle  qp \rangle ^\text{(ABOBA)} \\ \end{array}  \right ] = \left [ \begin{array}{c} K^{-1} \beta^{-1} \\ M \beta^{-1} \left(1-\frac{\delta t^2 K}{4M}\right)^{-1} \\ 0 \\ \end{array}  \right ],
\]
surprisingly giving exact values for the configurational average. Both schemes yield the same friction-independent upper-bound on the step size of $\delta t_\text{max}=2\sqrt{M/K}$: the (determinisitic) Verlet step size threshold. This implies that we can choose any timestep below this limit and still achieve perfect sampling of $\langle q^2 \rangle $, up to sampling error. This behavior is atypical of Langevin dynamics algorithms, for example comparing the  Bussi/Parrinello and BBK schemes we find
\[
\left [ \begin{array}{c} \langle  q^2 \rangle ^\text{(BP)} \\ \langle  p^2 \rangle ^\text{(BP)} \\  \langle  qp \rangle ^\text{(BP)}  \\ \end{array}  \right ] = \left [ \begin{array}{c} K^{-1} \beta^{-1} \left(1-\frac{\delta t^2 K}{4M}\right)^{-1} \\ M \beta^{-1} \\ 0 \\ \end{array}  \right ], \qquad \left [ \begin{array}{c} \langle  q^2 \rangle ^\text{(BBK)} \\ \langle  p^2 \rangle ^\text{(BBK)}  \\  \langle  qp \rangle ^\text{(BBK)} \\ \end{array}  \right ] = \left [ \begin{array}{c} K^{-1} \beta^{-1} \left(1-\frac{\delta t^2 K}{4M}\right)^{-1} \\ M \beta^{-1} \left(1+\frac{\gamma \delta t}{2}\right)^{-1} \\ 0 \\ \end{array}  \right ],
\]
giving identical second-order errors in configurational averages for this system, with the same value of $\delta t_\text{max}.$ The BBK scheme has a first order error in $\langle  p^2 \rangle $ that is also friction-dependent.

\begin{figure}[t]
  \centering
\includegraphics[width=5in]{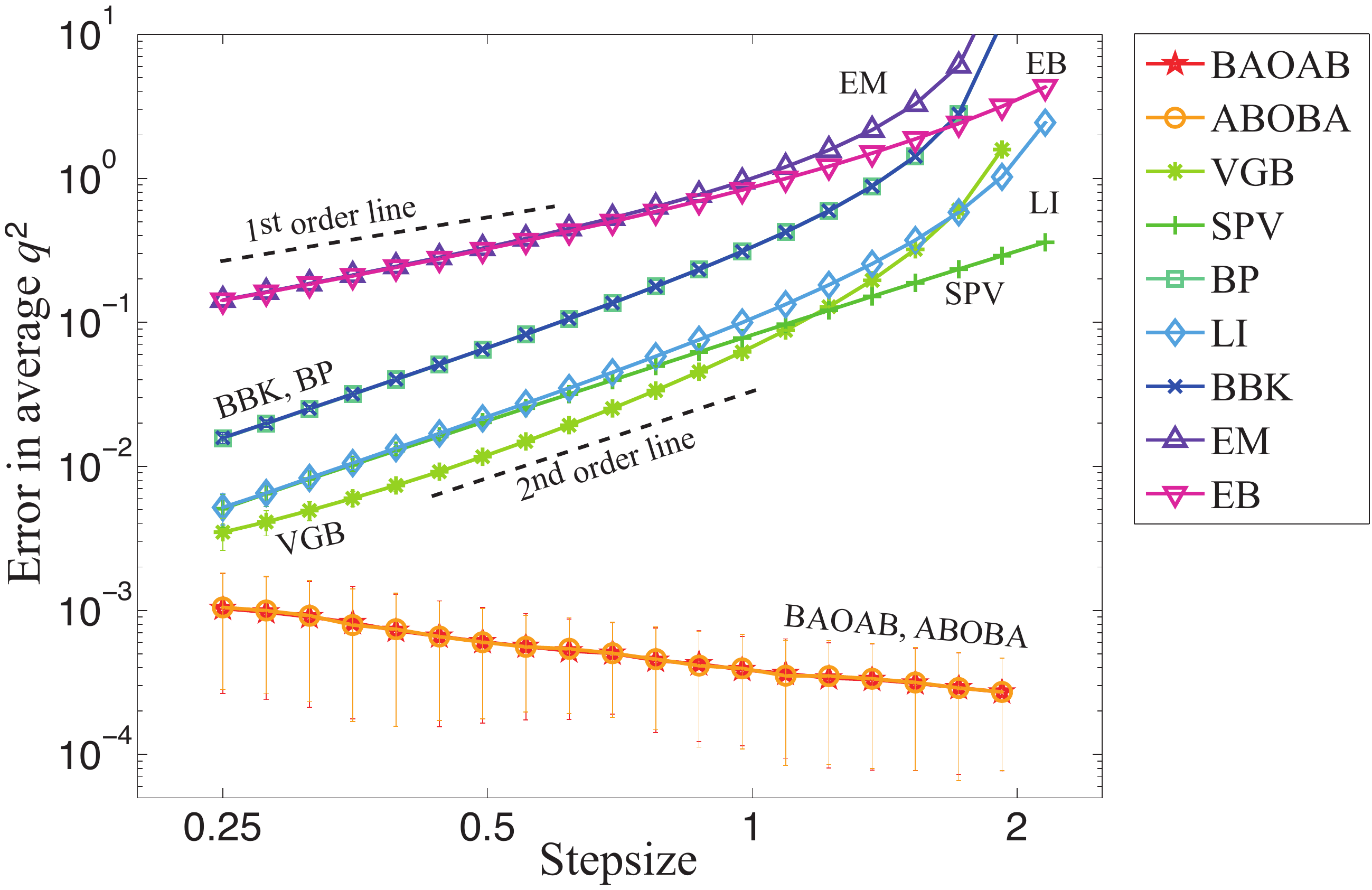}
\caption{The numerically-computed average error in $\langle  q^2 \rangle $, using the 1D harmonic oscillator with a given Langevin Dynamics method. Computation was fixed at $10^7$ total force evaluations, with $M=\gamma=\beta=K=1$. The results are averaged over 2000 independent repeat runs, with error bars included to give the standard deviation in these results. The exactness property of the ABOBA and BAOAB schemes result in the error decreasing as step size increases due to sampling error. \label{hopic}}
\end{figure}

The computed configurational averages for each scheme are shown in Table \ref{hotable}, while Figure \ref{hopic} shows the result of computing the value of $\langle  q^2 \rangle $ numerically, using a fixed total number of steps and varying the step size.   Three distinct regimes can be seen: the first-order ``Ermak'' methods, second-order methods and the exact methods (where any error comes solely from sampling error, rather than discretization error). 


We find that the method of van Gunsteren and Berendsen \cite{vgb88} is in fact 2nd order accurate for configurational sampling, not 3rd order as reported by those authors; we attribute this to a different notion of accuracy being used in that article.   

\begin{table}[bt]
  \centering
\begin{tabular}{|l|c||l|c|}
\hline
Scheme & $\langle q^2\rangle $ & Scheme & $\langle q^2\rangle $ \\ \hline
Exact & $K^{-1} \beta^{-1}$ & SPV & $K^{-1} \beta^{-1} \left(\gamma \, \delta t  \frac{1 - e^{-2\gamma  \delta t} }{2 \left(1 - e^{-\gamma  \delta t} \right)^{2} } \right)$ \\ \hline
BAOAB & $K^{-1} \beta^{-1}$ & LI & $K^{-1} \beta^{-1} - \frac{\delta t^2}{12 M \beta} +  O\left(\delta t^4 \right) $ \\ \hline
ABOBA & $K^{-1} \beta^{-1}$ & VGB & $K^{-1} \beta^{-1} + \frac{\gamma^2M - 2K}{24 M \beta K}\delta t^2 +  O\left(\delta t^4 \right) $ \\ \hline
BBK & $K^{-1} \beta^{-1} \left(1-\frac{\delta t^2 K}{4M}\right)^{-1}$ & EM & $K^{-1} \beta^{-1} \left(1-\frac{\delta t \, K}{2\gamma M}\right)^{-1}$ \\ \hline
BP & $K^{-1} \beta^{-1} \left(1-\frac{\delta t^2 K}{4M}\right)^{-1}$ & EB & $K^{-1} \beta^{-1} + \frac{\delta t}{2 \gamma M \beta} +  O\left(\delta t^2 \right) $ \\ \hline
\end{tabular}
\caption{The expected long-time computed average of $q^2$ using each Langevin dynamics scheme, for the 1D harmonic oscillator $U(q)=K q^2/2$. For brevity, some results are shown as leading order series in $\delta t.$ \label{hotable}}
\end{table}

\section{Error analysis for general systems }
Repeating the analysis in Section III for a more general $U(q)$ in a higher-dimensional setting is a challenging task for complicated schemes, but we do have a recently developed framework for carrying out the calculations.   As we have noted previously  we may define the invariant distribution of a second-order method $\hat{\rho}$ as the solution of the partial differential equation
\[
\hat{{\cal L}}^* \hat{\rho} = 0.
\]
Expanding the operator in a perturbation series in terms of the step size $\delta t$ and  combining equations (\ref{rhohat}) and (\ref{lhat}) yields
\[
 \left({{\cal L}}^*_{\rm LD} + \delta t^2 \hat{{\cal L}}^*_2 + \ldots \right) \rho_\beta \left(1  - \delta t^2 \beta f_2 + \ldots \right) = 0.
\]
Equating powers of the step size, we see that the order $0$ terms match automatically, leaving the leading order perturbation equation to be
\begin{equation}
 {{\cal L}}^*_{\rm LD} \left( \rho_\beta \, f_2 \right) = \beta^{-1} \hat{{\cal L}}^*_2 \rho_\beta. \label{perteqn}
\end{equation}
By the hypoelliptic property of the exact operator\cite{hormander}, the unique solution to ${{\cal L}}^*_{\rm LD} \phi =0$ is $\phi \propto \rho_\beta$, hence the homogenous solution to (\ref{perteqn}) is simply $f_2 = c$, a constant. Therefore we need only find a particular solution $f_2(q,p)$ solving (\ref{perteqn}) in order to find the leading order error in the long-time distribution $\hat{\rho}$. Once the perturbation is known, averages may be rebiased accordingly\cite{BoLe2007}, in effect increasing the order of the method. Of course $f_2$ may itself be a costly function to evaluate, involving a combination of high order derivatives of $U(q)$, and in the general case this is likely to lead to an inefficient method.

For the ABOBA and BAOAB methods, the right hand side of (\ref{perteqn}) is
\begin{align*}
- \beta^{-1} \hat{{\cal L}}^{*({\rm ABOBA})}_2 \rho_\beta &=  \frac{\gamma \rho_\beta}{4\beta} \left( \Delta_q M^{-1} U(q) - \beta p^T M^{-2} U''(q) p\right) + \frac{\rho_\beta}{4} p^T M^{-2} U''(q) \nabla_q U(q)\\
& \quad - \frac{\rho_\beta}{24} p^T \nabla_q p^T U''(q) M^{-3} p,\\
-\beta^{-1} \hat{{\cal L}}^{*({\rm BAOAB})}_2 \rho_\beta &= -\frac{\gamma \rho_\beta}{4\beta} \left( \Delta_q M^{-1} U(q) - \beta p^T M^{-2} U''(q) p\right) - \frac{\rho_\beta}{4} p^T M^{-2} U''(q) \nabla_q U(q)\\
& \quad + \frac{\rho_\beta}{12} p^T \nabla_q p^T U''(q) M^{-3} p ,
\end{align*}
where $U''(q) := \nabla_q \nabla_q^T U(q)$ is the hessian matrix of mixed partial derivatives.

Being able to write down equation (\ref{perteqn}) relies on the calculation of $\hat{{\cal L}}^*_2$, which involves the computation of a scheme's perturbed operator (\ref{lhat}), characterizing the evolution of a density of points in the phase space. This can be a challenge in itself, though for methods that derive from splitting the vector field, this operator is easily computed\cite{LeMa2012} using successive applications of the Baker-Campbell-Hausdorff (BCH) formula for products of exponentials \cite{BCH}.

As an example, consider the stochastic position verlet (SPV) method, using splitting pieces defined in equation (\ref{split2}). The infinitesimal generator associated to each part in (\ref{split2}) is given by
\[
{\cal L}^*_{A} \phi = -M^{-1} p \cdot \nabla_q \phi,\hspace{0.2in}
{\cal L}^*_{S} \phi = \nabla_q U(q) \cdot \nabla_p \phi + \gamma \nabla_p \left(\phi p \right) + \frac{\sigma^2}{2} M \Delta_p \phi.
\]
Note that the exact operator ${\cal L}^*_{\rm LD} = {\cal L}^*_{A} + {\cal L}^*_{S}.$ Using the notation from Section II, we code this numerical method as ``ASA''. The characteristic evolution operator $ {\cal L}^{*(SPV)}$ is then computed using
\[
 \exp\left( \delta t {\cal L}^{*(SPV)} \right) =  \exp\left( \left(\delta t/2\right) \, {\cal L}^{*}_A \right) \exp\left( {\delta t} {\cal L}^{*}_S \right) \exp\left( \left(\delta t/2\right) \, {\cal L}^{*}_A \right) ,
\]
where the BCH formula can be used to simplify the products of exponentials:
\[
\exp(t {\cal L}_1^*) \exp(t {\cal L}_2^*) = \exp\left( t\left({\cal L}_1^* + {\cal L}_2^*\right) + \frac{t^2}{2} \left[ {\cal L}_1^*,{\cal L}_2^* \right] + \frac{t^3}{12} \left([{\cal L}_1^*,[{\cal L}_1^*,{\cal L}_2^*]] - [{\cal L}_2^*,[{\cal L}_2^*,{\cal L}_1^*]]  \right) + O(t^4) \right),
\]
and $[{\cal L}_1^*,{\cal L}_2^*] = {\cal L}_1^*{\cal L}_2^* - {\cal L}_2^*{\cal L}_1^*$ is the commutator of ${\cal L}_1^*$ and ${\cal L}_2^*$.


By splitting the vector field (\ref{lang-1}-\ref{lang-2}), and choosing a preferred integration sequence, one can easily create and analyse a multitude of Langevin dynamics splitting methods using this technique; though it is perhaps surprising how small and subtle changes to the order of each piece's integration can yield vastly different average behavior in the long-time limit. This effect is most easily apparent if an asymmetry is created when two adjacent letters are swapped in a symmetric method. Symmetry ensures that the order of a method is at least two (by the Jacobi identity), while destroying this property could hamper stability as well as the order of the method.

Once the right hand side of (\ref{perteqn}) has been computed, solving to find the invariant density is an involving task and we do not pursue this here for the methods described above.  In the case of the ABOBA and BAOAB methods, solutions can be obtained as {\em doubly asymptotic expansions} in both $\delta t$ and the reciprocal friction coefficient $\gamma^{-1}$; moreover a superconvergence property can be demonstrated for BAOAB  configurational averages implying 4th order accuracy\cite{LeMa2012}. It is interesting to note that in the case of ABOBA, no such cancellation occurs, even though the methods have right hand sides that are apparently similar in the leading term.

The fundamental limitation of the asymptotic approach is that it remains to determine in which regime the theoretically obtained features of the perturbed distribution are manifest in simulation.   Large friction coefficient is known to reduce sampling efficiency, so we would need to work with modest values of $\gamma$, potentially invalidating the superconvergence property.  Likewise the crucial issue in many cases is the size of the allowable timestep for simulation, not the asymptotic error behavior for small step size.   These complexities must be addressed using computer experiment.

\section{Numerical results }
%

One of the most important features of a numerical method for ergodic dynamics (such as Langevin dynamics) is its preservation of the theoretical global phase space exploration rate. The spectral properties of the operator ${\cal L}^*_\text{LD}$ guarantee that we will explore the entire phase space (ergodicity), while the relatively small perturbations to the operator induced by numerical discretization are hoped not to significantly alter the rate of search. Ultimately, pushing the timestep up is the only way to breach timescale gaps, although this comes at the cost of corruption to the long-time averages.

The self-diffusion coefficient gives a metric quantifying the diffusion rate. It is often used as a way to compare the rate of phase space exploration between methods, and typically calculated using the integral of the velocity auto-correlation function. However, arbitrary methods can be constructed to artificially scale the velocity auto-correlation function, hence giving inaccurate diffusion constants.

Indeed, calculating the temperature of the system from an average of kinetic energy by
\begin{equation}
3N k_BT = \mathbb{E} \left[ p^T  M^{-1} p \right], \label{kinetictemp}
\end{equation}
gives a similar problem.  Alternative functions, including functions of $q$ only, can be obtained whose averages are proportional to the system temperature\cite{Rugh,LanLif,JepAytEva}. Such ``configurational temperature" observables are normally based on the periodic forces of the system in order to work in periodic boundary conditions; in the droplet simulations reported below (i.e. without boundary conditions) we used the simpler expression:
\begin{equation}
 3 N k_BT = \mathbb{E} \left[ q \cdot \nabla U(q)\right].\label{configtemp}
\end{equation}
 Were one able to solve the dynamics exactly, the kinetic and configurational temperatures would of course be equal. However, using a numerical method in the large-timestep regime we instead sample expectations with respect to the perturbed density $\hat{\rho}$, that may introduce discrepancies between configurational and kinetic temperatures.   It is our view that a configuration-based temperature calculation is normally more useful and relevant for assessing the quality of configurational sampling methods.

In a similar way, the speed of exploration of the space should not be determined solely from functions of momentum, but should rely on actual barrier crossing rates or times to reach some target region of phase space.

\subsection{One-dimensional double well}
The advantages of performing tests initially on a simple model are that  (i) the exact solution is known (or can be numerically integrated to arbitrary precision),  while (ii) the model's simplicity allows us to perform exhaustive computation to refine results and determine asymptotic properties. Here we use the algorithms to integrate Langevin dynamics for  a one-dimensional model with potential function $U(q)=(q^2-1)^2+q$, a double-well. This example is well-studied as an approximation for modelling a dual state system. We use unit mass, friction and temperature, and test a range of step sizes, beginning at $\delta t=0.2$ and increasing by $5\%$ until we reach a step size where all of the methods are no longer stable. We run $500$ independent experiments for each step size, with computation fixed at $10^9$ iterations for each realisation.

\begin{figure}[hb]
\centering
\includegraphics[width=\textwidth]{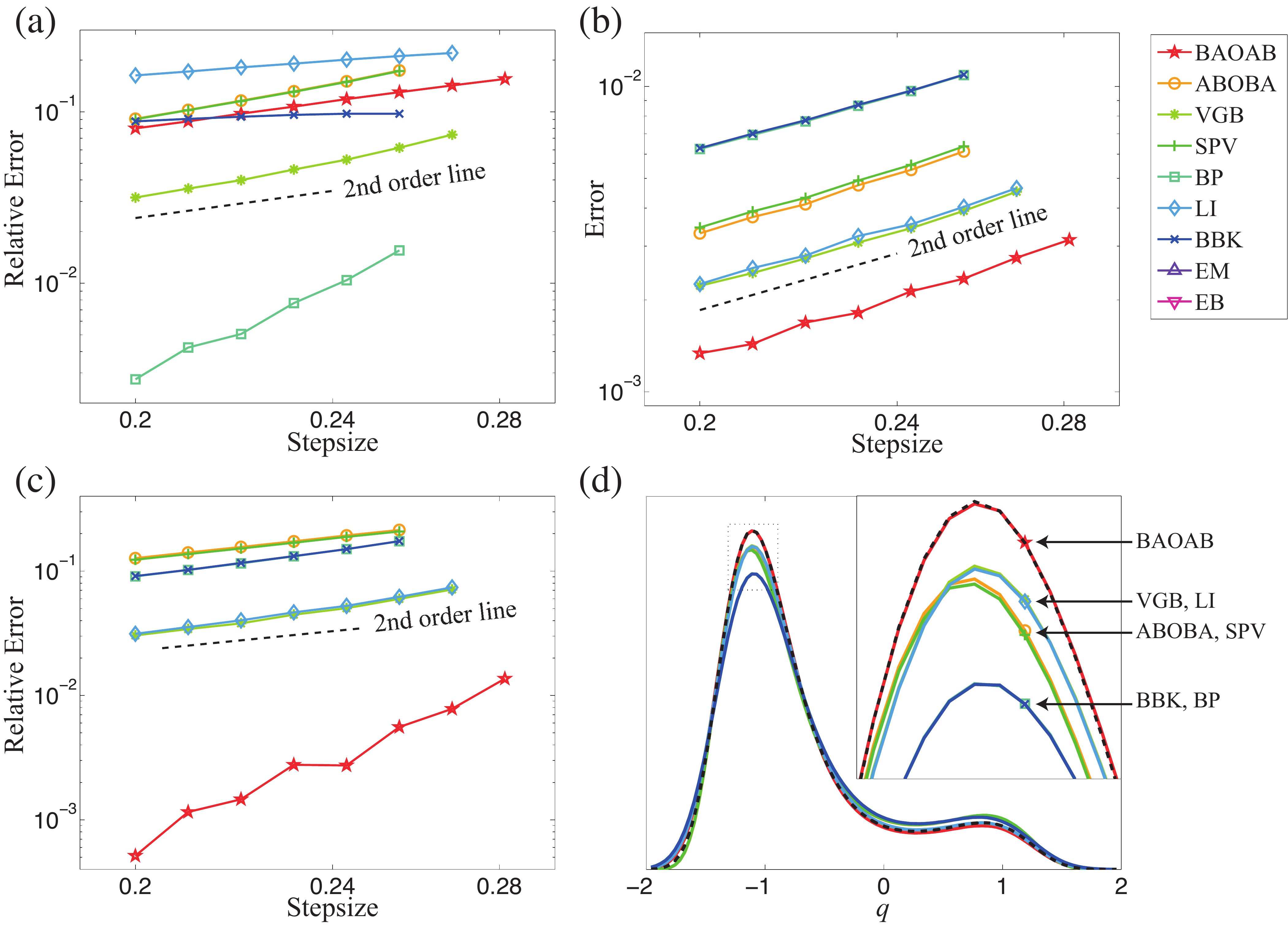} 
\caption{The BAOAB scheme is shown to significantly reduce configurational discretization errors, which distort averages in Langevin Dynamics, for the  one-dimensional double-well system with potential energy function $U(q)=(q^2-1)^2+q$. Errors are computed from the average of 500 independent trajectories, with $10^9$ total iterations per trajectory. The step sizes tested began at $\delta t=0.2$ and were increased by $5\%$ incrementally until all schemes became unstable. The relative error in the temperature (computed by averaging the momenta) is given in (a), with the scheme of Bussi/Parrinello giving a high order relationship with the step size. This is contrasted in (c) where the temperature is calculated using the instantaneous system position; here it is instead the BAOAB scheme that gives a high order result. The error in configurational distribution (calculated as a root mean squared deviance in the  histogram bins) is shown in (b),   with a sample computed distribution given in (d) for $\delta t=0.25.$ The exact distribution is shown as a dashed line, with the inset magnifying the density of the deepest well.}
  \label{dwell}
\end{figure}

The error in configurational distribution  is estimated by dividing the interval $[-2,2]$ into 16 equal bins, and calculating the observed configurational density in each bin for every computed trajectory. The error in the observed densities for each bin are calculated by comparing the absolute difference between the observed and exact expected densities (the latter obtained using a high-order numerical solver). The overall error in the configurational density is calculated from the root mean squared value of these errors. 
Additionally, we calculate the observed kinetic temperature (\ref{kinetictemp}) and configurational temperature (\ref{configtemp}) for each method. The results are shown in Figure \ref{dwell}.

For the configurational distribution, all the methods shown give a second-order relation in the step size, in contrast to Figure \ref{hopic}. Notably, the BAOAB and ABOBA methods are no longer exact for this anharmonic model, while the two first-order methods do not appear at all, as neither of them is stable in this region. Of the methods that are stable, the BAOAB method gives both the largest usable timestep and the smallest maximum error in the configurational distribution for any given timestep.  
A sample plot of the computed distribution for all schemes at $\delta t=0.25$ is also given in Figure  \ref{dwell}d, it is clear from this that the BAOAB scheme performs exceptionally well.

Of particular note is the apparent lack of direct correspondence between the errors in configurational temperature and kinetic temperature. The Bussi/Parrinello scheme is shown to preserve the kinetic temperature to a very high degree, with less than a $1\%$ error for $\delta t<0.25,$ while, at the same step size, the BAOAB integrator gives more than $10\%$ discrepancy in kinetic temperature. However, these results are inverted when looking at configurational sampling accuracy (and configurational temperature).    Clearly, if maintaining the configurational averages is the goal,  estimating the fidelity of the calculation by relying on the kinetic temperature is a risky strategy.   A much more reliable approach is to make use
of the configurational temperature, although even this does not give the complete story, since in the example at hand the configurational temperature scales with a high power of the step size, while configurational sampling error declines as the second power of $\delta t$.  The second order behavior does not contradict our previous observations\cite{LeMa2012} as we are here far from the large $\gamma$ limit, but does indicate that the superconvergence property is not the key feature at play in the setting of this model problem.

\subsection{Alanine dipeptide}
In our next experiments, we studied the alanine dipeptide molecule, a classic test case for molecular dynamics. We compare computed averages for solvated and unsolvated alanine dipeptide using the BAOAB, ABOBA, van Gunsteren and Berendsen, Bussi and Parrinello, Langevin Impulse, Stochastic Position Verlet (SPV) and the Br\"{u}nger/Brooks/Karplus (BBK) schemes.  We obtained poor results using the first-order schemes and therefore did not consider them here.   To provide a means of calculating basline values, we use the stochastic position verlet (SPV) method with a small step size, for which the discretization error is essentially negligible.

We implement each of the methods in the NAMD lite package\cite{namdlite2007}, and observe the effect of discretization error (if any) on computed  configurational averages. The CHARMM22 forcefield was used to compute force interactions.

\subsubsection{Unsolvated}
We simulate the alanine dipeptide molecule (22 atoms) in vacuum at 300K for 2.5ns for multiple different step sizes and friction constants, to observe how different simulation parameters affect computed averages. Parameters for each run were taken from a $50\times50$ grid, with each point on the grid corresponding to a $(\delta t, \gamma)$ parameter set for a simulation. The parameters for the bottom-left point on this grid are $\delta t_1=1$fs, $\gamma_1 = 10^{-2}/$ps, where each grid point moving upward gives a $20.7\%$ increase in the friction value used, while each grid point moving right gives a $2.46\%$ increase in step size. These ratios were chosen so as to give a broad range of parameter sets to test over, while ensuring that the range was not so wide as to yield a large number of unsuitable parameter sets (for example, using an unstable step size) leading to wasted computation. All the schemes were unstable for the maximum step size tested ($\delta t_{50}=3.29$fs).

\begin{figure}[hp]
\centering
 \includegraphics[clip=true,trim=0in 0.425in 0in 0in]{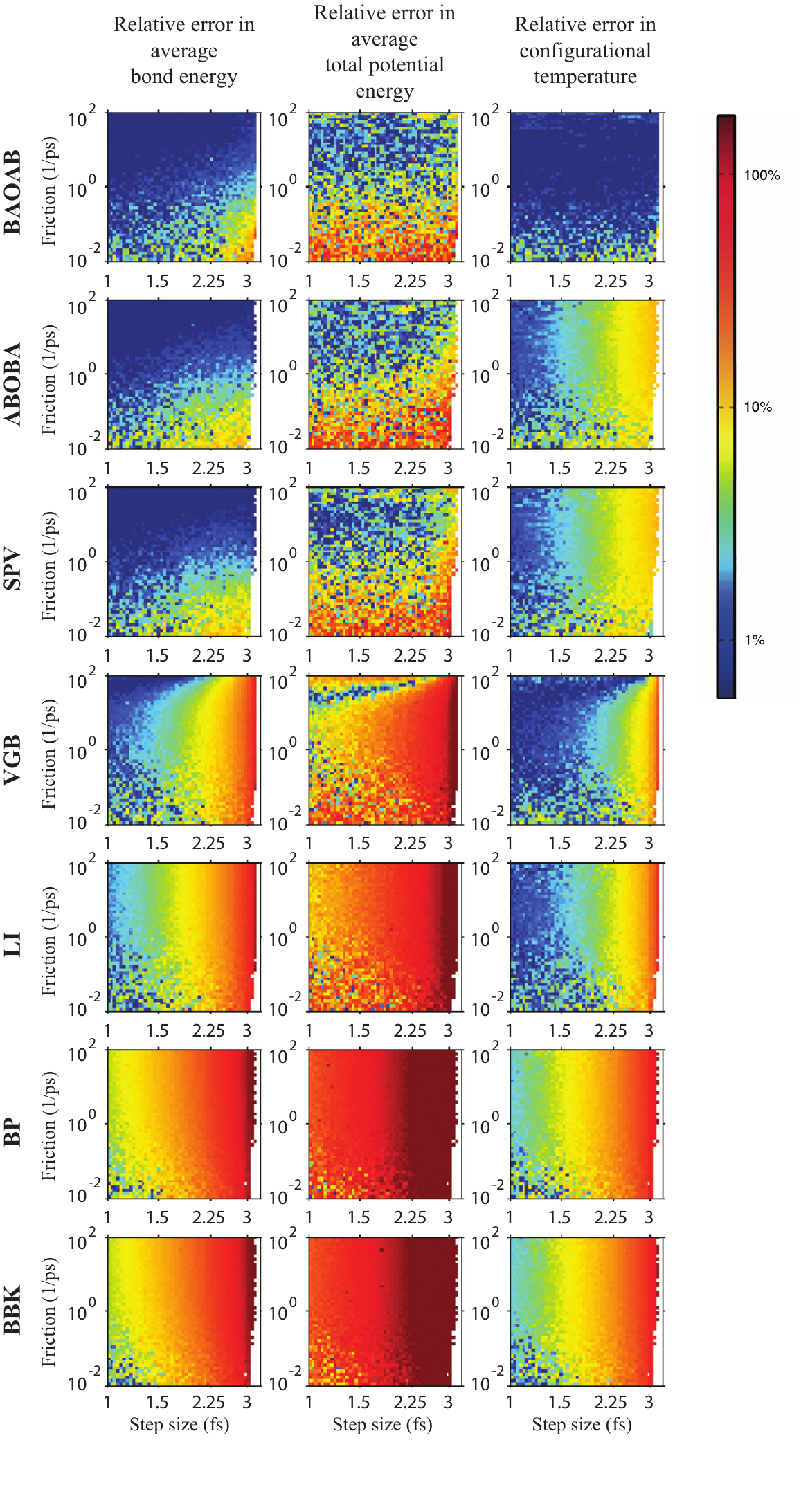}
 \caption{Results from $2.5$ns simulations of alanine dipeptide in vacuum at the given step size (horizontal) and friction (vertical). Pixels are colored according to relative errors for each simulation, with white pixels indicating instability. \label{unsolv}}
\end{figure}

The results of the simulations for each scheme are given in Figure \ref{unsolv}, where we color points on the $50\times50$ grid of parameter sets to indicate the results from that respective simulation. Relative errors are calculated in the average total potential energy and the average total bond energies, where the ``exact'' comparison value is taken from averaging ten 2.5ns runs using the SPV scheme at $\delta t=0.25$fs, where it is expected that discretization error is not significant.

With such a small simulation we would perhaps expect to see a very ``noisy'' result: high variances due to the sampling error vastly outweighing the discretization error. But in fact the discretization error dominates and is observable at step sizes significantly below the stability threshold.

White pixels in the grids in Figure \ref{unsolv} represent a method's instability, showing that in general there is a small stability threshold increase for the large-friction case. 
There is no significant increase in this threshold between the methods however, with BAOAB, VGB and LI schemes giving a marginal increase over the others.

One salient feature of the results of Figure \ref{unsolv}, is that for the BAOAB scheme there is consistently less than a $1\%$ error in the computed configurational temperature (for moderate friction) across all step sizes, even the largest stable timesteps tested. The relative errors obtained were so small that no discernable trend (with step size) can be shown, due to the sampling error, whereas the other schemes tested show an error consistent with second-order schemes (an example is given in Appendix \ref{app_secondorder}). 

\begin{figure}[ht]
\centering
 \includegraphics[width=\textwidth,clip=false]{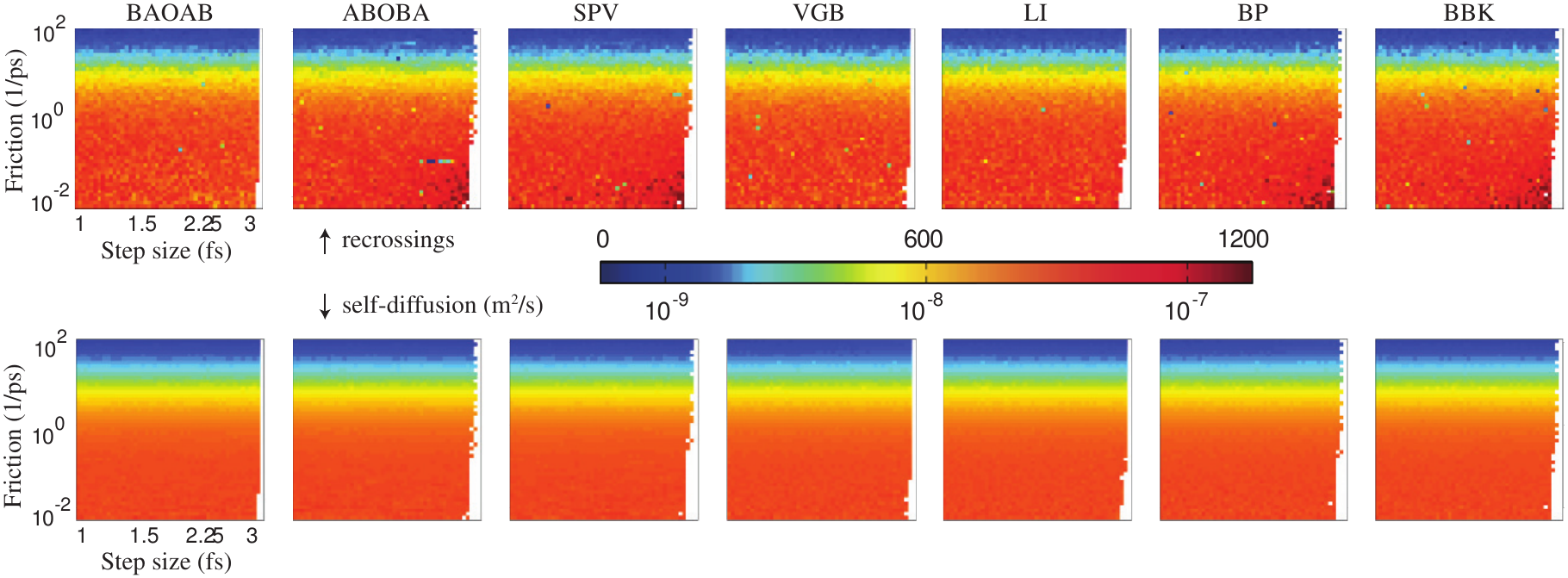}
\caption{The number of barrier recrossings (top) and the diffusion coefficients (bottom) are shown for each simulation, the latter in $\text{m}^2/$s and computed by integrating the velocity autocorrelation function over an interval of 1ps. As expected, the computed coefficients do not vary significantly between methods, though changing the friction above 1/ps has a substantial effect.
}
  \label{diffcoeff}
\end{figure}

Self-diffusion coefficients are calculated from integrating the computed velocity autocorrelation function, where a history is kept of the velocities for 1ps. The values plotted in Figure \ref{diffcoeff} show that changing the step size within the indicated range using any of the schemes has only a very slight effect on the diffusion coefficient, while increasing the friction can dramatically reduce it.   Examining the graphs, we settle on $\gamma=1/$ps as the largest value of $\gamma$ for which the diffusion coefficient is unperturbed for all the schemes.  It is interesting that larger damping parameters do not substantially improve numerical stability for any of the methods, except in an extreme case for the VGB method ($\gamma\approx 100/$ps, where the diffusion constant is drastically reduced).

Numerical values for the computed average potential energy are given in Table \ref{adptable} for varying step size at $\gamma = 1/$ps.

\begin{table}[ht]
  \centering
\begin{tabular}{|l|c|c|c|c||c|c|c|c|}
\hline
\multirow{2}{*}{Scheme} & \multicolumn{4}{|c||}{\shortstack{Average total potential energy\\(kcal/mol)}} &  \multicolumn{4}{|c|}{\shortstack{Total number of observed\\recrossings}} \\ \cline{2-9}
& $\delta t=1.5$fs & $\delta t=2$fs & $\delta t=2.5$fs & $\delta t=3$fs & $\delta t=1.5$fs & $\delta t=2$fs & $\delta t=2.5$fs & $\delta t=3$fs \\ \hline
BAOAB & $1.65 \pm 0.04$ & $1.67 \pm 0.05$ & $1.68 \pm 0.03$ & $1.70 \pm 0.05$ & $821 \pm 17$ & $814 \pm 25$ & $823 \pm 21$ & $798 \pm 11$ \\ \hline
ABOBA & $1.69 \pm 0.05$ & $1.70 \pm 0.03$ & $1.75 \pm 0.03$ & $1.87 \pm 0.04$ & $823 \pm 17$ & $829 \pm 20$ & $826 \pm 23$ & $833 \pm 26$ \\ \hline
SPV & $1.70 \pm 0.13$ & $1.70 \pm 0.05$ & $1.74 \pm 0.06$ & $1.88 \pm 0.03$ & $811 \pm 56$ & $822 \pm 22$ & $838 \pm 26$ & $835 \pm 22$ \\ \hline
VGB & $1.89 \pm 0.03$  & $2.14 \pm 0.05$ & $2.65 \pm 0.06$ & $4.27 \pm 0.04$ & $802 \pm 19$ & $804 \pm 28$ & $812 \pm 16$ & $813 \pm 17$ \\ \hline
LI & $2.05 \pm 0.05$ & $2.42 \pm 0.04$ & $3.21 \pm 0.06$ & $5.84 \pm 0.06$ & $837 \pm 33$ & $819 \pm 19$ & $822 \pm 24$ & $821 \pm 27$ \\ \hline
BP & $2.75 \pm 0.05$  & $3.91 \pm 0.06$ & $6.19 \pm 0.05$ & $14.0 \pm 0.18$ & $828 \pm 27$ & $801 \pm 19$ & $821 \pm 21$ & $818 \pm 41$ \\ \hline
BBK & $2.78 \pm 0.03$ & $3.89 \pm 0.05$ & $6.21 \pm 0.06$ & $13.9 \pm 0.10$ & $826 \pm 18$ & $825 \pm 24$ & $834 \pm 19$ & $835 \pm 16$ \\ \hline \hline
\emph{Baseline} & \multicolumn{4}{|c||}{ $1.66 \pm 0.04$} & \multicolumn{4}{|c|}{ $808 \pm 28$ }\\ \hline
\end{tabular}
\caption{Numerical results for ten $2.5$ns simulations of unsolvated alanine dipeptide, with friction set to \mbox{$\gamma = 1/$ps}. The mean and standard deviations of all simulations are given. The baseline comparison run was completed by averaging ten $2.5$ns simulations using $\delta t=0.25$fs with the SPV scheme. Sampling error will play a large role in the determination of these averages, but it is clear that the BAOAB scheme outperforms the others by a significant margin. The number of observed recrossings is obtained by counting the number of times the central dihedral angles in the alanine dipeptide model hop between their two configurations. \label{adptable}}
\end{table}

If we accept, say, a  5\% error tolerance for the average potential energy, we see that BAOAB admits a usable step size of up to $3$fs, whereas the ABOBA and SPV schemes are restricted to a neighborhood of $2$fs, with the usable timestep threshold for other methods well below $1.5$fs.


\subsubsection{Solvated}
We immerse the alanine dipeptide molecule in a sphere of TIP3P water (10A radius, total system is 424 atoms) and equilibrate for 1ns at 300K to generate an initial configuration. We then run simulations using each scheme considered in the unsolvated case, using a 10A cutoff for electrostatics and van der Waals potentials. The value of friction was fixed at $1/$ps, with runs performed with increasing step size. Initial timesteps were $\delta t = 2$fs, with subsequent simulations increasing the step size by $5\%$, until reaching a step size where all of the methods fail. Each simulation was performed for $5$ns at $T=300$K using spherical (harmonically restrained) boundary conditions.  Although the particular boundary conditions may not be representative of all biomolecular simulations, we contend that the crucial features of numerical stability and relative method performance are unaffected by the particular choice.

\begin{figure}[hbt]
\centering
\includegraphics[width=\textwidth]{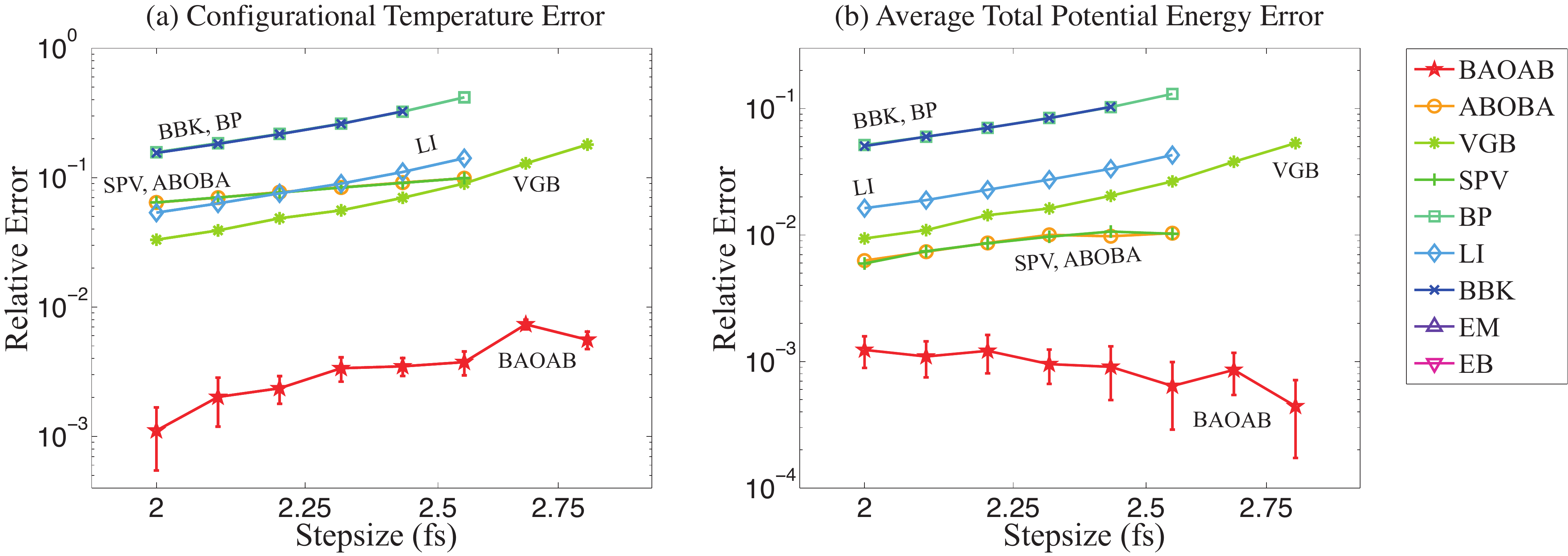}
\caption{Numerical results from 5ns runs of alanine dipeptide solvated in a 10A sphere of TIP3P water are shown, using the given algorithms. Errors are computed against a baseline solution averaged from ten 5ns simulations using the SPV scheme at $\delta t=0.5$fs. Results from a single run are shown for each scheme except in the case of the BAOAB method.   The BAOAB scheme shows an order of magnitude improvement in the error in computed average total potential energy; because of the small absolute errors, we exhibit the means and standard deviations from 10 runs for each step size used.   The lack of an observed trend line for BAOAB suggests that the discretization error is being dominated by the sampling error.}
  \label{solv}
\end{figure}

The results are given in Figure \ref{solv}. Compared to the scheme of Bussi and Parrinello, the relative error in average total potential energy using the BAOAB scheme is seen to be smaller by two orders of magnitude when computed at a step size around $\delta t=2.5$fs. The surprising downward trend of the error using the BAOAB scheme could be indicative of higher-order terms dominating in the error expansion, showing that our asymptotic approach does not give definitive answers about bevavior in the large step size regime. The analytic results obtained for the discretization error are understood only for \mbox{$\delta t \rightarrow 0$}.  More detailed analytical investigation of this phenomenon is beyond the scope of this article.

The breakdown of results for all energy contributions is given in Appendix \ref{app_solvated}. It is clear from these results that the average bond energy is a crucial component in explaining the results of  Figure \ref{solv}. The average bond energy computed using the BAOAB scheme gives a flat profile with respect to the timestep increasing, whereas many other methods demonstrate an extreme drift approaching the stability threshold, causing a large error in the average total potential energy. The averages of other energies do not significantly contribute to the observed errors.

The contribution of error coming from the restraining boundary condition energy was extremely small, suggesting that the properties of the bulk water in the model are responsible for the differences in efficiency seen here. Hence we would expect the obseved corruption of averages to be generalizable to any simulations involving other boundary conditions, or other simulations involving water.

\FloatBarrier

\section{Conclusion }
We have studied a total of nine different integration methods for the Langevin dynamics equations, including popular schemes that are in widespread use for molecular sampling.   We have seen that some of these can be derived as splitting methods and in a few cases the perturbation of the invariant distribution has been determined in some regime (for example, the small step size, large friction limits).  It is also possible to solve for the error in averages as a function of step size in the case of a harmonic oscillator, which we believe has direct relevance for biomolecular modelling where the bond stretches are modelled as harmonic restraints.  Harmonic models are also likely to relate well to simulations of crystalline materials\cite{SmMa1995}.   Our analyses show that a particular ordering of the building blocks of a splitting method, the BAOAB integrator\cite{LeMa2012}, provides exact configurational averages for the harmonic oscillator and  4th order accurate configurational averages for a general nonlinear model in the large friction limit.   We have examined the performance of this method in relation to other schemes for toy models and for small biomolecular models both with and without solvent, with the observation that the analytical results on the error in distribution are highly correlated to their performance in practice.  In particular, the BAOAB method performs very differently than the other methods in practical simulations, giving much higher accuracies (particularly for the configurational temperature) up to the Verlet stability threshold.

A surprising observation is that discretization error, not sampling error, dominates in the simulations we performed, which involved a common small biomolecular test system and a time interval of only a few nanoseconds.

Let us put the numerical results into perspective.  Our simulations explore only a few of the available quantities that might be relevant for modelling.  It is interesting that all of the second order schemes tested provided reasonable transition rates (in terms of barrier crossings), and so would give a similar rate of exploration of the phase space.  Since molecular dynamics is often used for phase space exploration and supplemented by other techniques for precise averaging, the methods may have some utility regardless of the fact that they provide in some cases very poor approximation of  configurational averages.  It is also possible for a scheme to accurately resolve one quantity but not another (for example, in the case of unsolvated alanine dipeptide, the scheme of van Gunsteren and Berendsen gives reasonably good configurational temperatures but poor average potential energies).

The ABOBA and SPV methods perform very similarly, and reasonably well, both for energy calculations and in terms of configurational temperature in all of the numerical experiments performed with alanine dipeptide.  The similarity between these methods is a consequence of the fact that both are drift-kick-drift style algorithms, with a subtle difference in their ``kick'' updates: the ABOBA scheme solves (\ref{lang-2}) in a leapfrog manner by splitting off the force from the Ornstein-Uhlenbeck stochastic term, where as the SPV scheme solves equation (\ref{lang-2}) exactly for constant position $q$. One may expect that an exact solve would provide the method better properties, but in practice the leapfrog splitting in ABOBA is advantageous in the high friction regime. For large $\gamma$, the result of solving exactly means that the  momentum update becomes dominated by the noise, shrinking the contribution from the force term. The advantage of splitting up (\ref{lang-2}) is that this isolates the force from the noise, integrating it separately, making ABOBA (and indeed other schemes using the same splitting strategy, such as BAOAB and Bussi and Parrinello) effective for any value of $\gamma \ge 0$.

Using quantities based on the momenta to estimate temperature or diffusion constants is called into question; certainly the connection between the accuracy of the kinetic energy average and the accuracy of other more directly relevant quantities is weak.    The kinetic temperature measure is an accurate approximation of the true temperature in the case of the VGB scheme, but this same method gives relatively poor potential energy averages.   

We conclude by emphasizing that for bond energy and total potential energy averages, in vacuum simulation, the BAOAB method performs better than the other methods and is substantially better at large step sizes, giving a larger useful range of step size by a factor of at least 25\% with an order of magnitude smaller errors at large step size.  The differences are magnified still further when configurational temperatures are compared.



%
%

%

\begin{acknowledgments}
We thank David Hardy (University of Illinois) for his support with the modification of the NAMD package. We also appreciate the support of the Lorentz Center (Leiden, NL) and the programme on ``Modelling the Dynamics of Complex Molecular Systems'' which supported the authors and provided valuable interactions during the preparation of the article.  This work has made use of the resources provided by the Edinburgh Compute and Data Facility (http://www.ecdf.ed.ac.uk/). The ECDF is partially supported by the eDIKT initiative (http://www.edikt.org.uk).   We further acknowledge the support of the Engineering and Physical Sciences Research Council which has funded this work as part of the Numerical Algorithms and Intelligent Software Centre under Grant EP/G036136/1.
\end{acknowledgments}


\nocite{*}
\providecommand{\noopsort}[1]{}\providecommand{\singleletter}[1]{#1}%

\newpage

\appendix

\section{Implementation details of Figure 2} \label{app_fig2}

We consider a single particle confined to the plane, with instantaneous horizontal and vertical position denoted $x,y \in \mathbb{R}$ respectively. The particle feels a force with respect to the potential energy function
\[
U(x,y) = \frac{1 + 5\Big(1 - \exp\Big(-250\left(y-0.25 \right)^2 \Big) \Big) \, \exp\Big(-100 x^2 \Big)}{ \left(\sqrt{x^2+y^2}-1 \right)^2} + \frac{4 \, \exp\Big( -20 x^2 \Big) }{5} ,
\]
qualitatively giving an energy surface with a narrow transition pathway between two basins. We seek to sample the canonical distribution $\bar{\rho}_\beta(x,y)$ using Langevin dynamics, comparing the computed distributions given by the BAOAB and Bussi/Parrinello algorithms (given in the following section) at varying stepsizes.

Figure 2 corresponds to numerical experiments using fixed friction constant $\gamma=10$ and temperature $\beta=1$. From left to right, the stepsizes tested were \mbox{$\delta t=  \left[0.005,0.021,0.024,0.027,0.03 \right]$}. Each image is a two dimensional histogram over the unit square centered on the origin, with 100 equally spaced bins in both directions. The computed density shown for each timestep is averaged from 64 runs of $10^8$ steps.

\section{Numerical methods} \label{app_methods}
We present the numerical methods used in this article, assuming timestep $\delta t$, friction constant $\gamma$ and temperature $T$ with diagonal mass matrix $M$ and position and momentum vectors $q,p$ respectively. The force is $F(q) := -\nabla U(q)$ and $k_B$ is Boltzmann's constant. $R_n$ is a $3N$-vector of independent, identically distributed normal random numbers with zero mean and unit variance. Where $J>1$ random numbers are required per degree of freedom, multiple independent (uncorrelated) random vectors are denoted $R^{(j)}_n, j=1,\ldots,J$ in the schemes.

\newpage

\noindent {\bf BAOAB} \newline
\noindent Note: This scheme is available in recent versions of NAMD (after Jan 2013) by including options \mbox{`\emph{langevin  on}'} and \mbox{`\emph{langevinBAOAB  on}'} in the input parameter file.
\begin{align*}
p_{n+1/3} &= p_n + \frac{\delta t}{2} F(q_n), \\
q_{n+1/2} &= q_n + \frac{\delta t}{2} M^{-1} p_{n+1/3}, \\
p_{n+2/3} &= e^{-\gamma \delta t} p_{n+1/3} + \sqrt{ k_BT \left( 1 - e^{-2\gamma \delta t} \right) } M^{1/2} R_n, \\
q_{n+1} &= q_{n+1/2} + \frac{\delta t}{2} M^{-1} p_{n+2/3}, \\
p_{n+1} &= p_{n+2/3} + \frac{\delta t}{2} F(q_{n+1})
\end{align*}
{\bf ABOBA}
\begin{align*}
q_{n+1/2} &= q_n + \frac{\delta t}{2} M^{-1} p_{n}, \\
p_{n+1/3} &= p_n + \frac{\delta t}{2} F(q_{n+1/2}), \\
p_{n+2/3} &= e^{-\gamma \delta t} p_{n+1/3} + \sqrt{ k_BT \left( 1 - e^{-2\gamma \delta t} \right) } M^{1/2} R_n, \\
p_{n+1} &= p_{n+2/3} + \frac{\delta t}{2} F(q_{n+1/2}),\\
q_{n+1} &= q_{n+1/2} + \frac{\delta t}{2} M^{-1} p_{n+1}
\end{align*}
{\bf Van Gunsteren/Berendsen (VGB)}\newline
\noindent We must initialize the vector $X$,
\[
X_1 = \kappa_4 M^{-1/2} R^{(3)}_0,
\]
and then iterate
\begin{align*}
V_{n+1} &= \kappa_1 M^{-1/2} R^{(1)}_n, \\
\hat{V}_{n+1} &= \kappa_2 X_n + \kappa_3 M^{-1/2} R^{(2)}_n, \\
p_{n+1} &= e^{-\gamma \delta t} p_n + \frac{1-e^{-\gamma \delta t} }{\gamma} F(q_n) + M \left( V_{n+1} - e^{-\gamma \delta t} \hat{V}_{n+1} \right), \\
X_{n+1} &= \kappa_4 M^{-1/2} R^{(3)}_n, \\
\hat{X}_{n+1} &= \kappa_5 V_{n+1} + \kappa_6 M^{-1/2} R^{(4)}_n, \\
q_{n+1} &= \frac{e^{\gamma \delta t/2} - e^{-\gamma \delta t/2}}{\gamma} M^{-1}p_{n+1} + X_{n+1} - \hat{X}_{n+1},
\end{align*}
where we use $3N$--vectors $X,\hat{X},V,\hat{V}.$ Constants $\kappa_i$ are given as
\begin{align*}
\kappa_1 &= \sqrt{k_BT \left( 1 - e^{-\gamma \delta t} \right) }, \\
\kappa_2 &= \frac{2\gamma - \gamma e^{\gamma \delta t/2} - \gamma e^{-\gamma \delta t/2}}{\gamma \delta t - 3 + e^{-\gamma \delta t} \left(4e^{\gamma \delta t/2} - 1 \right)}, \\
\kappa_3 &= \sqrt{k_BT} \sqrt{ \frac{\gamma \delta t \left( e^{\gamma \delta t} -1 \right) - 4\left(e^{\gamma \delta t /2} -1 \right)^2 }{\gamma \delta t - 3 + e^{-\gamma \delta t} \left(4e^{\gamma \delta t/2} - 1 \right)} }, \\
\kappa_4 &= \gamma^{-1} \sqrt{k_BT} \sqrt{\gamma \delta t - 3 + e^{-\gamma \delta t} \left(4e^{\gamma \delta t/2} - 1 \right) },\\
\kappa_5 &= \gamma^{-1} \left( \frac{2 - e^{\gamma \delta t} - e^{-\gamma \delta t} }{e^{-2\gamma \delta t} -1} \right), \\
\kappa_6 &= \gamma^{-1} \sqrt{k_BT} \sqrt{ \frac{\gamma\delta t \left( e^{-\gamma \delta t} -1 \right) + 4\left( e^{-\gamma \delta t/2} -1 \right)^2 }{e^{-\gamma \delta t} - 1} }.
\end{align*}
{\bf Stochastic Position Verlet (SPV)}
\begin{align*}
q_{n+1/2} &= q_n + \frac{\delta t}{2} M^{-1} p_{n}, \\
p_{n+1} &= e^{-\gamma \delta t} p_{n} + \frac{1-e^{-\gamma \delta t}}{\gamma} F(q_{n+1/2}) + \sqrt{ k_BT \left( 1 - e^{-2\gamma \delta t} \right) } M^{1/2} R_n, \\
q_{n+1} &= q_{n+1/2} + \frac{\delta t}{2} M^{-1} p_{n+1}
\end{align*}
{\bf Bussi/Parrinello (BP)}
\begin{align*}
p_{n+1/4} &= e^{-\gamma \delta t/2} p_{n} + \sqrt{ k_BT \left( 1 - e^{-\gamma \delta t} \right) } M^{1/2} R^{(1)}_n, \\
p_{n+2/4} &= p_{n+1/4} + \frac{\delta t}{2} F(q_{n}), \\
q_{n+1} &= q_n + {\delta t} M^{-1} p_{n+2/4}, \\
p_{n+3/4} &= p_{n+2/4} + \frac{\delta t}{2} F(q_{n+1}), \\
p_{n+1} &= e^{-\gamma \delta t/2} p_{n+3/4} + \sqrt{ k_BT \left( 1 - e^{-\gamma \delta t} \right) } M^{1/2} R^{(2)}_n
\end{align*}
{\bf Langevin Impulse (LI)} \newline
\noindent We use the algorithm designed for configurational sampling; a correction term is given in \cite{LI} to improve the sampling of momenta, though this has no effect on configurational averages. We must initialize the $3N$--vector $Z$,
\[
Z_1 = M^{1/2} \left(\alpha_0 R_0 + \hat{\alpha} R_1 \right).
\]
and then iterate
\begin{align*}
p_{n+1/4} &= p_n + \omega \delta t F(q_n), \\
p_{n+2/4} &= e^{-\gamma \delta t/2} \left(p_{n+1/4} + \omega Z_n  \right), \\
q_{n+1} &= q_n + \frac{1 - e^{-\gamma \delta t}}{\gamma e^{-\gamma \delta t/2}} M^{-1} p_{n+2/4}, \\
Z_{n+1} &= M^{1/2} \left( \alpha R_n + \hat{\alpha} R_{n+1} \right), \\
p_{n+3/4} &= e^{-\gamma \delta t/2} p_{n+2/4} + \hat{\omega}Z_{n+1} , \\
p_{n+1} &= p_{n+3/4} + \hat{\omega}F(q_{n+1}),
\end{align*}
where
\begin{gather*}
\omega = \frac{e^{-\gamma \delta t} + \gamma \delta t - 1 }{\gamma \delta t \left( 1 - e^{-\gamma \delta t} \right)}, \qquad \hat{\omega} = 1 - \omega, \\
a = k_BT \left(2 \omega^2 \gamma \delta t + \omega - \hat{\omega} \right),\\
b = k_BT \left(2 \omega\hat{\omega} \gamma \delta t + \hat{\omega} - \omega \right),\\
c = k_BT \left(2 \hat{\omega}^2 \gamma \delta t + \omega - \hat{\omega} \right),\\
\alpha = 2^{-1/2} \sqrt{c + a + \sqrt{(c+a)^2 - 4b^2}},\\
\hat{\alpha} = 2^{-1/2} \sqrt{c + a - \sqrt{(c+a)^2 - 4b^2}}, \\
\alpha_0 = \sqrt{\alpha^2 - c}.
\end{gather*}

\noindent {\bf Br\"{u}nger/Brooks/Karplus (BBK)}
\begin{align*}
p_{n+1/2} &= \left(1 - \frac{\gamma \delta t}{2} \right)  p_n + \frac{\delta t}{2} F(q_{n}) + \frac{1}{2} \sqrt{2\gamma k_BT \delta t} M^{1/2} {R_n}, \\
q_{n+1} &= q_{n} + {\delta t} M^{-1} p_{n+1/2}, \\
p_{n+1} &= \left(1 + \frac{\gamma \delta t}{2} \right)^{-1} \left( p_{n+1/2} + \frac{\delta t}{2} F(q_{n+1}) + \frac{1}{2}\sqrt{2\gamma k_BT \delta t} M^{1/2} R_{n+1} \right),
\end{align*}
{\bf Ermak/McCammon (EM)}

\noindent As we consider only the scalar friction case, the update scheme for the position reduces to the Euler-Maruyama algorithm, with rescaled timestep.
The update scheme for the momenta is unused in our numerical experiments, but can be found in \cite{ErMc}. We iterate
\begin{align*}
q_{n+1} = q_n + \frac{\delta t D}{k_BT} M^{-1}  F(q_n) + \sqrt{2 D \delta t } M^{-1/2} R_n,
\end{align*}
where
\[
D = k_BT / \gamma.
\]

\noindent {\bf Ermak/Buckholtz (EB)}
\begin{align*}
p_{n+1} &= e^{-\gamma \delta t} p_n + \frac{1-e^{-\gamma \delta t}}{\gamma} F(q_n) + \sqrt{ k_BT \left( 1 - e^{-\gamma \delta t} \right) } M^{1/2} R^{(1)}_n, \\
q_{n+1/3} &= q_n + M^{-1}\left(p_n + p_{n+1} - 2\gamma^{-1}F(q_n)  \right) \frac{1 - e^{-\gamma \delta t}}{\gamma \left( 1 + e^{-\gamma \delta t} \right) }, \\
q_{n+2/3} &= q_{n+1/3} + \gamma^{-1} \delta t M^{-1} F(q_n), \\
q_{n+1} &= q_{n+2/3} + \gamma^{-1} \sqrt{2k_BT \left(\gamma \delta t - 2\frac{1 - e^{-\gamma \delta t}}{1 + e^{-\gamma \delta t} } \right)}M^{-1/2} R^{(2)}_n
\end{align*}

\section{Second order behavior of schemes} \label{app_secondorder}
We demonstrate the results for the relative error in configurational temperature, for simulations of alanine dipeptide in a vacuum at fixed friction $\gamma = 1.94/$fs. This is equivalent to plotting a horizontal cross-section of the results given in Figure 5, at the corresponding friction value. We find that in all but one scheme a clear second order trend is visible; the exception is BAOAB which has much higher accuracy than the other methods and for which a trend line could not be resolved; we conjecture that for BAOAB the order of accuracy of the configurational temperature is substantially higher than two (consistent with our observations for one degree-of-freedom anharmonic cases).
\begin{figure}[hb]
\centering
\includegraphics[scale=0.4]{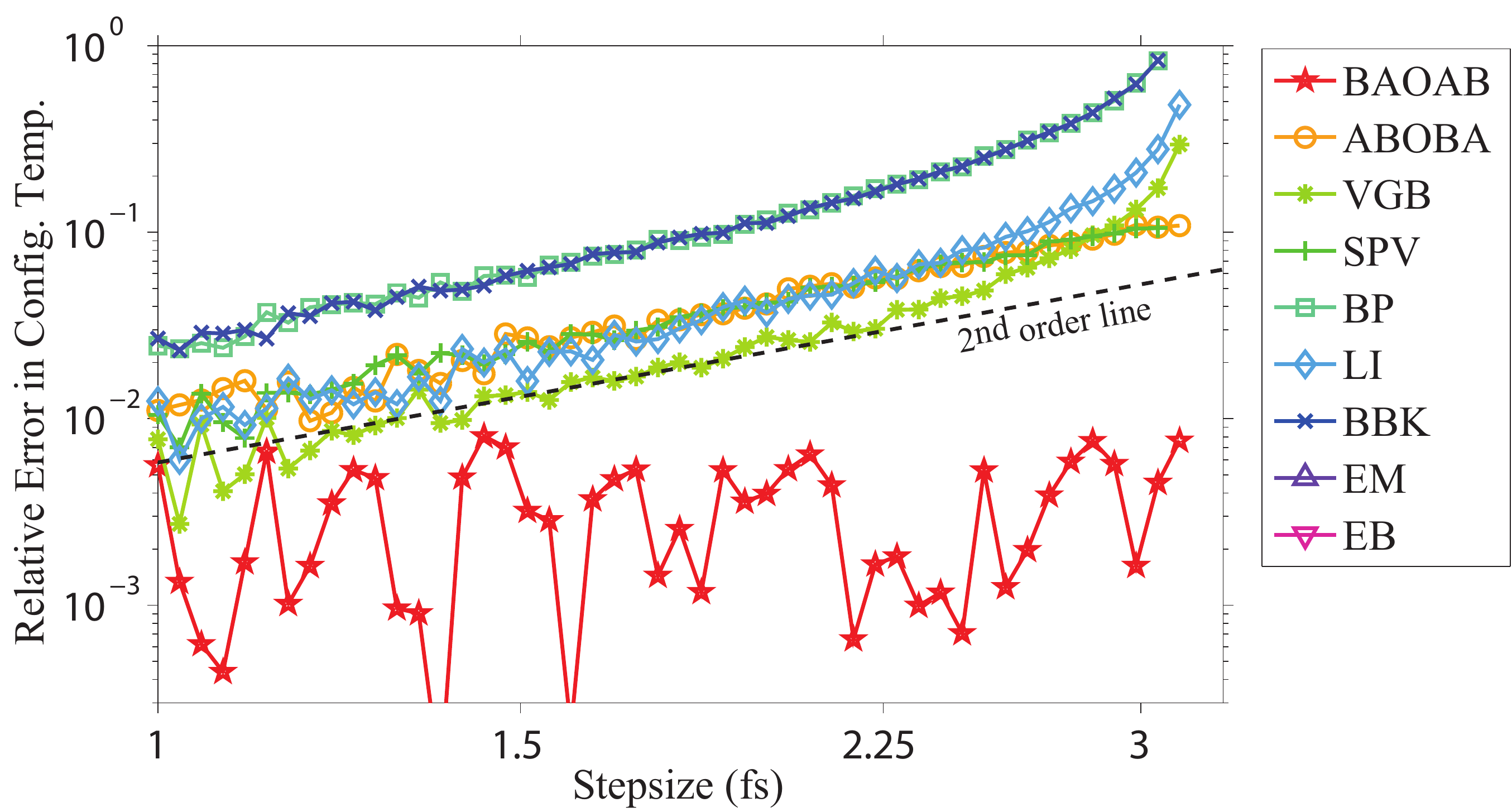}
\end{figure}
\section{Solvated results}\label{app_solvated}
The breakdown of computed average energies are given for each method. The black dashed line marks the baseline solution for comparison.
\begin{figure}[hb]
\centering
\includegraphics[width=\textwidth]{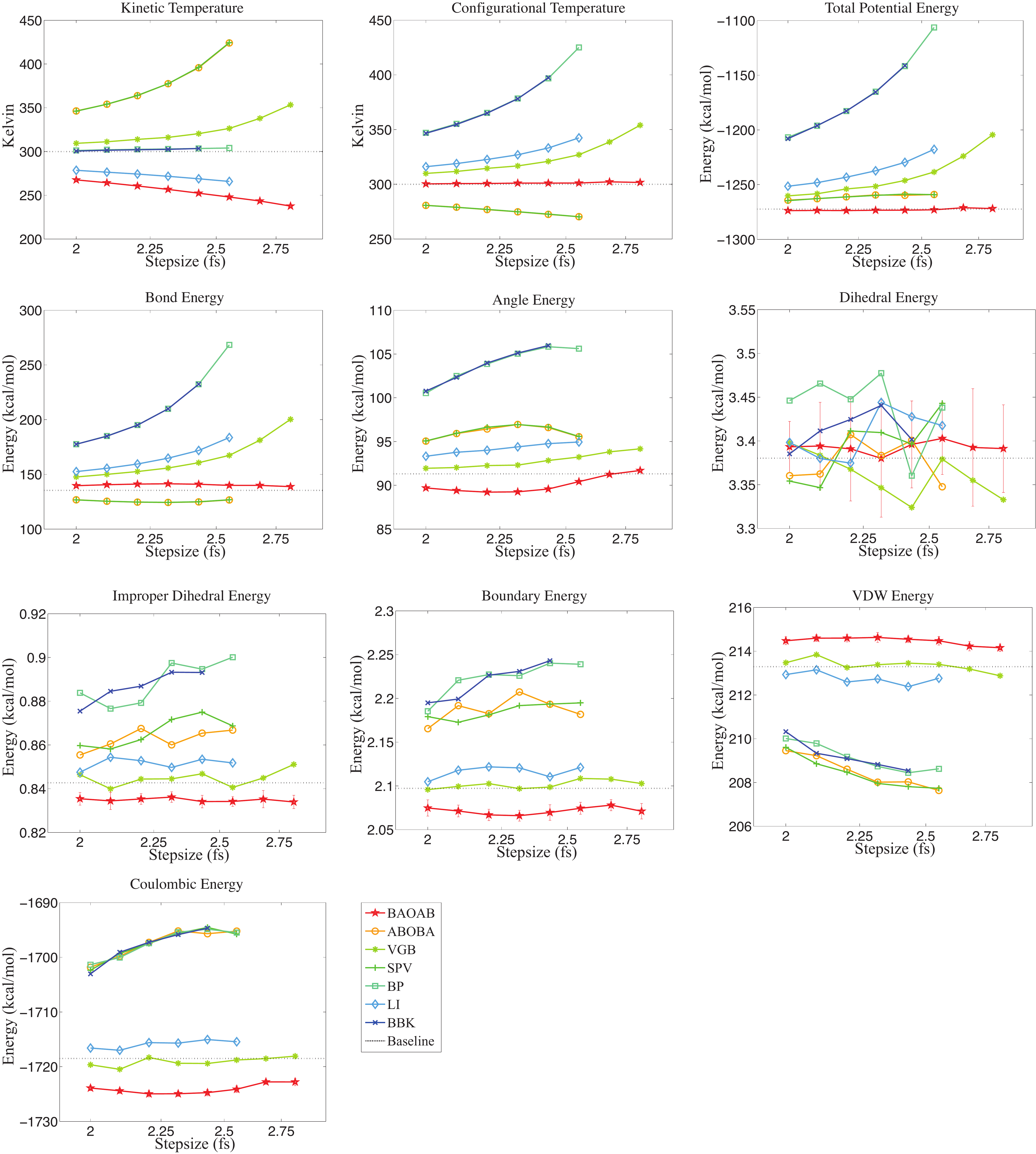}
\end{figure}

\end{document}